# Enabling Adaptive Rate and Relay Selection for 802.11 Mobile Ad Hoc Networks

Neil Mehta, Alexandra Duel-Hallen, and Wenye Wang

Department of Electrical and Computer Engineering,

North Carolina State University,

Raleigh, NC 27695, USA

nbmehta2@ncsu.edu, sasha@ncsu.edu, wwang@ncsu.edu

*Abstract*—Mobile ad hoc networks (MANETs) are self-configuring wireless networks that lack permanent infrastructure and are formed among mobile nodes on demand. Rapid node mobility results in dramatic channel variation, or fading, that degrades MANET performance. Employing channel state information (CSI) at the transmitter can improve the throughput of routing and medium access control (MAC) protocols for mobile ad hoc networks. Several routing algorithms in the literature explicitly incorporate the fading signal strength into the routing metric, thus selecting the routes with strong channel conditions. While these studies show that adaptation to the time-variant channel gain is beneficial in MANETs, they do not address the effect of the outdated fading CSI at the transmitter. For realistic mobile node speeds, the channel gain is rapidly varying, and becomes quickly outdated due the feedback delay. We analyze the link throughput of joint rate adaptation and adaptive relay selection in the presence of imperfect CSI. Moreover, for an 802.11 network that employs geographic opportunistic routing with adaptive rate and relay selection, we propose a novel method to reduce the effect of the feedback delay at the MAC layer in the presence of Rayleigh fading. This method exploits channel reciprocity and fading prediction and does not require significant modification to the existing 802.11 frame structure. Extensive network simulations demonstrate that the proposed approach significantly improves the throughput, delay, and packet delivery ratio for high mobile velocities relative to previously proposed approaches that employ outdated CSI at the transmitter.

## I. INTRODUCTION

Rapid node mobility in MANETs can result in dramatic channel variation, or fading, that degrades link lifetime and network throughput [1]. Routing and MAC protocols that exploit fading channel-state-information (CSI) at the transmitter have the potential to improve upon conventional routing protocols and have attracted the attention of researchers, e.g. [2- 7]. These methods explicitly incorporate the fading signal strength to select the route or next hop with strong channel conditions. Combined adaptive routing and adaptive rate selection was investigated in [3-6]. While these studies show that adapting to the time-variant channel gain is beneficial in MANETs, most do not focus on the effect of outdated CSI at the transmitter.

Channel-aware routing and MAC approaches can be viewed as a subclass of adaptive transmission techniques where the transmitted signal varies with fading and multiple access interference. To realize the potential of adaptive transmission, the transmitter needs accurate CSI for the upcoming transmission frame. The CSI is usually estimated at the receiver and fed back to the transmitter. However, unless the mobile speed is very low, the estimated CSI cannot be used directly to select the parameters of adaptive transmission systems since it quickly becomes outdated due to the rapid channel variation caused by multipath fading. To enable adaptive transmission for mobile radio systems, prediction of future fading channel samples has been investigated [8]. While fading prediction was extensively studied for cellular radio systems, it has only recently attracted the attention of researchers in the ad hoc networking literature, e.g. to improve the interface between the physical and the MAC layer [9- 11] and to enable channel-aware routing [12]. However, the prediction algorithms implemented or assumed in these papers are based on the predictors employed in the cellular radio research and are not realistic. While the accuracy/performance/complexity trade-offs of prediction-enabled adaptive transmission are well-understood in cellular systems, these results are not indicative of the throughput, delay, outage probability, packet delivery ratio, etc. of channel-adaptive MANET protocols. Fading prediction techniques for MANETs need to take into account the network topology, multiple network layers, moving receivers and transmitters, packet-based transmission, noise, and interference. Moreover, prediction errors, and imperfect CSI in general, should be taken into account in the protocol design and analysis.

In this paper, we investigate adaptive rate-and-relay selection for the IEEE 802.11 network with Greedy Perimeter Stateless Routing (GPRS) [13]. A progress-throughput product metric was employed in [4]. This metric favors next-hop relays with strong instantaneous link power and good progress to the destination. Significant throughput degradation was demonstrated in [4] for fast vehicular speeds due to the outdated CSI caused by the feedback delay from the candidate relay nodes. In this paper, we develop a method to reduce the CSI delay significantly by exploiting channel reciprocity possible in the IEEE 802.11 systems due to half-duplex transmission. The CSI error is further reduced using fading prediction. The effect of outdated CSI is analyzed analytically, and OmNet++ [14] simulations are employed to demonstrate the impact of the proposed CSI reduction on the network throughput and end-to-end delay.

This research was supported by the ARO grant W911NF-10-1-0394 and NSF grant CNS-1018447.

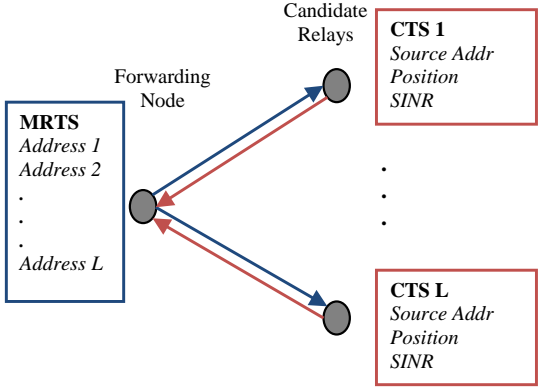

Figure 1: Polling current position and CSI [4].

## II. CHANNEL-ADAPTIVE RATE AND RELAY SELECTION

Our system is based on the adaptive rate-and-relay selection protocol described in [4]. It employs the IEEE 802.11b Distributed Coordination Function (DCF) at the MAC layer and Greedy Perimeter Stateless Routing (GPSR) protocol [13] at the routing layer. The routing protocol selects up to $L$ next-hop relay candidates by averaging the Signal-to-Interference-and-Noise Ratio (SINR) from the received HELLO messages of the neighboring nodes, broadcast at intervals of 1.5 seconds by all the nodes. The routing layer then passes the information of the next-hop relay candidates to the MAC layer, which transmits a Multiple Request-To-Send (MRTS) packet to the next-hop candidates, as illustrated in Figure 1. If this request is successfully received, and the relay is available to receive a packet, it responds with a clear-to-send (CTS) packet, which contains its position information and the received SINR. To avoid collisions, the responses are sent in the specified order, and the "best" relay is then selected by the transmitter.

To make relay selections at the routing and MAC layers, the progress-throughput product metric is computed as

$$\max_{l,i}\left\{Z_l \lambda_{i_l}(\gamma_l)\right\} \qquad (1)$$

where $i_l$ is the data rate for the relay $l=1\ldots L$ ($L$ is the total number of next-hop relay candidates), $\gamma_l$ is the SINR of the relay link $l$, $\lambda_{i_l}(\gamma_l)$ is the expected throughput to relay $l$ at data rate $i_l$, and $Z_l$ is the progress offered by to relay $l$ towards the destination. The above routing metric aims to achieve a balance by selecting routes that provide the best tradeoff between packet progress and link throughput.

## III. LINK PERFORMANCE ANALYSIS FOR IMPERFECT CSI

In a fixed rate system, one of the four available data rates in the IEEE 802.11b physical layer is employed, i.e. 1, 2, 5.5, and 11 Mbps. The average throughput for Rayleigh fading channels using data rate $i$ is then given by [4]

$$\lambda_i = \frac{1}{D_i}\int_0^\infty P_{s,i}(\gamma)f(\gamma)d\gamma \qquad (2)$$

where $P_{s,i}(\gamma)$ is the probability of successfully transmitting a packet at data rate $i$, $D_i$ is the duration of the transmission (in seconds) including the acknowledgement (ACK), and $f(\gamma)=\exp(-\gamma/\overline{\gamma})/\overline{\gamma}$ is the probability density function (pdf) of the SNR for the Rayleigh fading channel, where $\overline{\gamma}$ is the average SNR. The probability of a successful packet transmission can be approximated as

$$P_{s,i}(\gamma)=[1-P_{b,i}(\gamma)]^N \qquad (3)$$

where $P_{b,i}(\gamma)$ is the bit error probability at SNR $\gamma$ and data rate $i$, and $N$ is the total number of bits to be transmitted including ACK.

The performance of the rate-and-relay adaptive metric that maximizes the throughput in a system with $L$ relays was analyzed for ideal CSI in [4]. This analysis is extended to the imperfect CSI case below and provides insight into the performance of the progress-and-throughput product metric (1). Assuming ideal CSI, the link throughput of rate-and-relay adaptation is [4]

$$\lambda_{IDEAL} = \int_0^\infty \max_i\left\{\frac{P_{s,i}(\gamma)}{D_i}\right\}f(\gamma)d\gamma, \qquad (4)$$

where the probability density function of $\gamma$ (the maximum SNR among $L$ relays) is given by

$$f(\gamma)=\frac{L}{\overline{\gamma}}e^{-\gamma/\overline{\gamma}}(1-e^{-\gamma/\overline{\gamma}})^{L-1}.$$

While (4) assumes ideal CSI, to utilize adaptive rate and relay selection in practice, received SNR has to be fed back to the transmitter, and thus becomes outdated due to the feedback delay. Even with estimation and prediction, the outdated CSI is imperfect. This estimation error needs to be taken into account in the throughput computation. Assuming the Minimum Mean Square Error (MMSE) estimate of the actual SNR $\gamma$ given the outdated SNR $\hat{\gamma}$, the throughput of the rate-and-relay adaptive system is

$$\lambda(\hat{\gamma})=\max_i \frac{P_{s,i}(\hat{\gamma})}{D_i}. \qquad (5)$$

While the metric (5) is simple, it is sub-optimal since it does not employ the information about the estimation accuracy. The optimal metric is given by

$$\lambda_O(\hat{\gamma})=\max_i \frac{1}{D_i}\int_0^\infty P_{s,i}(\gamma)f(\gamma\mid\hat{\gamma})d\gamma, \qquad (6)$$

where the conditional pdf of $\gamma$ given $\hat{\gamma}$ is [15],

$$f(\gamma\mid\hat{\gamma})=\frac{1}{\overline{\gamma}(1-\rho)}I_0\left(\frac{2\sqrt{\rho\gamma\hat{\gamma}}}{\overline{\gamma}(1-\rho)}\right)\exp\left(\frac{-1}{\overline{\gamma}(1-\rho)}\left(\gamma+\hat{\gamma}\right)\right), \qquad (7)$$

where $I_0(\cdot)$ is the zero-th-order modified Bessel function, $0\le\rho\le1$ is the cross-correlation between $\gamma$ and $\hat{\gamma}$, and $\overline{\gamma}$ is the average symbol SNR. Assuming unit variance for the signal power, the normalized mean-square-error (NMSE) $\sigma^2$ is given by

$$\sigma^2=1-\rho \qquad (8)$$

The average optimal and sub-optimal throughputs using adaptive rate-and relay selection are

$$\lambda_O = \int_0^\infty \lambda_O(\hat{\gamma})f(\hat{\gamma})d\hat{\gamma} \quad \text{and} \quad \lambda = \int_0^\infty \lambda(\hat{\gamma})f(\hat{\gamma})d\hat{\gamma}, \qquad (9)$$

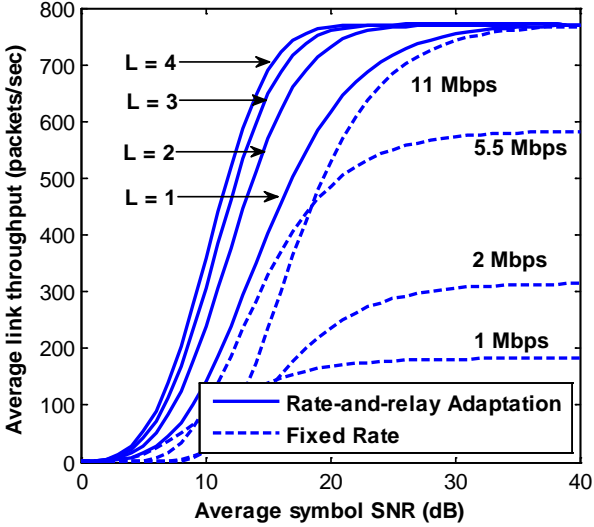

Figure 2: Average throughput vs. average symbol SNR. Fixed rate and rate-and-relay selection methods with perfect CSI. Rayleigh fading channel.

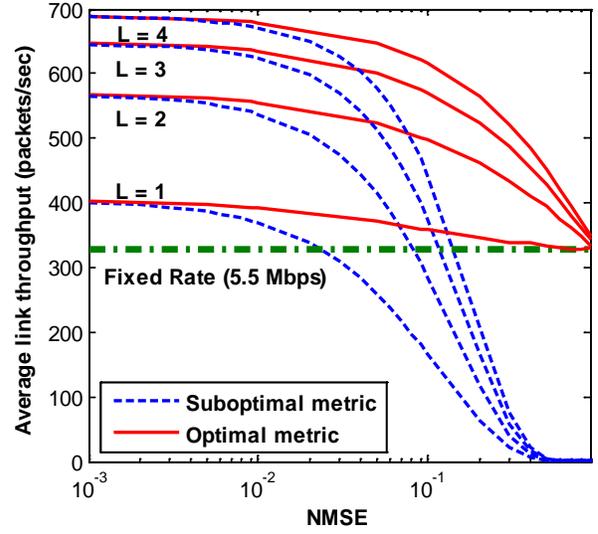

Figure 3: Average throughput vs. NMSE, rate-and-relay selection, Rayleigh fading with average symbol SNR = 15 dB.

respectively, where the density function of $\hat{\gamma}$ for $L^{th}$-order relay selection diversity is [15,**Error! Bookmark not defined.**]

$$f(\hat{\gamma}) = \frac{L}{\overline{\gamma}(1-\sigma^2)} e^{\frac{-\hat{\gamma}}{\overline{\gamma}(1-\sigma^2)}} \left(1 - e^{\frac{-\hat{\gamma}}{\overline{\gamma}(1-\sigma^2)}}\right)^{L-1} \quad (10)$$

Figure 2 shows the average link throughput (in packets/second) vs. the average symbol SNR for fixed rate systems (2) and channel-adaptive systems with ideal CSI (4). These results differ from those in [4] and were obtained using the Bit Error Rate (BER) approximations for the 802.11b modulation methods in [16,17]. In Figure 3, we plot the average throughput with imperfect CSI as a function of the NMSE for average symbol SNR = 15dB for the optimal and the suboptimal metrics (9). As the NMSE → 0, these equations tend to the throughput with the ideal CSI (4) shown in Figure 2. On the other hand, as the NMSE increases the correlation between $\gamma$ and $\hat{\gamma}$ decreases. As a result, $\lambda_O(\hat{\gamma})$ tends to $\max_i \lambda_i(\gamma)$, i.e. the maximum among the throughputs of fixed rate methods (2) for given SNR, (e.g. 5.5 Mbps for the average SNR = 15 dB in agreement with Figure 2). On the other hand, since $\hat{\gamma} \to 0$ as NMSE → 1, the suboptimal throughput $\lambda(\hat{\gamma})$ (5) tends to zero. We observe that the suboptimal metric has near-optimal performance for NMSE < 0.01, but degrades rapidly beyond this threshold. In the remainder of the paper we employ a simple suboptimal metric

$$\max_{l,i} \left\{ Z_l \lambda_{i_l}(\hat{\gamma}_l) \right\}$$

and focus on reducing the CSI error to achieve reliable transmission.

## IV. IMPROVED CSI QUALITY USING CHANNEL RECIPROCITY AND FADING PREDICTION

It was demonstrated in Section III that accurate CSI is essential to enable adaptive rate and relay selection. In this section, we focus on practical approaches for improving the quality of the CSI at the MAC layer. Consider the adaptive rate-and-relay selection protocol [4] at the MAC layer of 802.11 summarized in Section II. In this protocol, the next-hop relay candidates measure the SINR while receiving the MRTS packet and send it back to the sender in the CTS packet. As the CSI is obtained during reception of the MRTS packet, we refer to this scheme as RTS CSI. An example timeline of this MRTS-CTS exchange for $L = 4$ next-hop candidates is illustrated in Figure 4. The time delay in the measured CSI from the RTS packet to the actual CSI during transmission of data packet is $\tau_{MRTS} = 2$ms. Note that this delay denotes the time interval by which the channel measurement gets outdated. As $L$ increases, this delay increases causing degradation in performance for higher Doppler frequencies [4,7].

In this paper, we propose a novel method of CSI estimation at the transmitter that exploits the reciprocity in the channel to reduce the time delay by which the channel measurement gets outdated. Due to half-duplex transmission in the 802.11 networks, both directions of transmission utilize the same carrier frequency. Thus, the complex channel gain is identical at a given time at both ends of the wireless link. The sender can measure CSI for each potential relay link using the CTS signal. We term this method CTS CSI. As the CTS packets are transmitted sequentially, the delay $\tau_l$ for the $l^{th}$ CTS packet can be quantified as,

$$\tau_l = (L - l)(T_{SIFS} + T_{CTS}) + T_{SIFS} \quad (11)$$

where $T_{SIFS}$ is the duration of the short inter-frame space (SIFS) interval, and $T_{CTS}$ is the duration of the CTS packet.

As illustrated in Figure 4, employing channel reciprocity reduces the time delay between the measured CSI and actual CSI by estimating the SINR during reception of the CTS packet at the transmitter side with a peak delay of $\tau_1 = 1.5$ms for the first CTS packet as compared to RTS CSI wherein the SINR is estimated at the receiver side during reception of the MRTS packet with a delay of $\tau_{MRTS} = 2$ms.

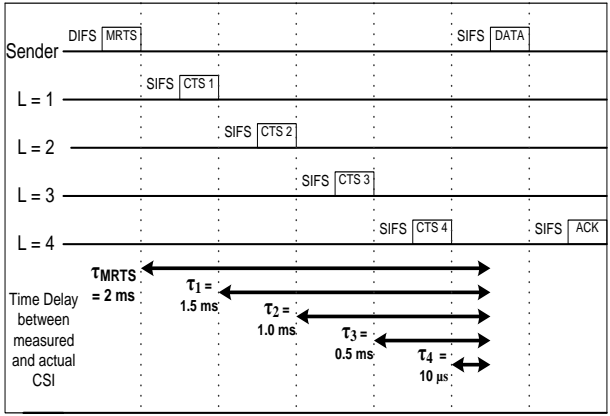

Figure 4: Time line of the MRTS-CTS exchange between the sender and four candidate relays, number of relays $L = 4$.

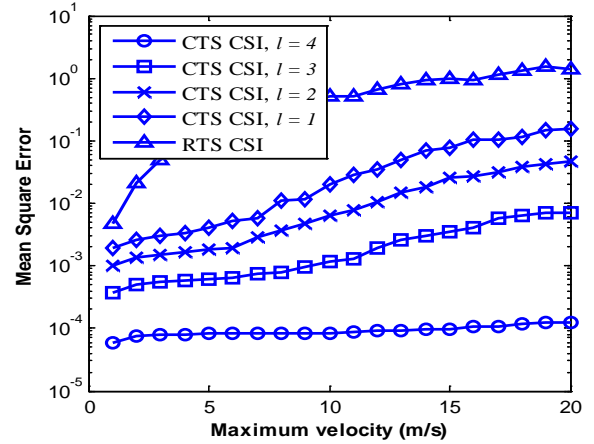

Figure 5: Comparison of Mean Square Error for the CTS CSI method with delays $\tau_l$ in Figure 4 and for RTS CSI, flat Rayleigh fading, average SNR = 30dB, total number of relays $L = 4$.

However, there is still a delay $\tau_l$ for the $l^{th}$ CTS packet between measuring the CSI and utilizing it for rate-and-relay selection when transmitting the data packet. We employ the MMSE complex fading channel prediction [8] during the reception of the $l^{th}$ CTS packet to predict the channel $\tau_l$ seconds ahead in time to compensate for this delay.

To realize the predictor, pilot symbols spaced 10 data symbols apart, are inserted in the CTS packet. Since the CTS packet contains 496 symbols at the 1 Mbps rate, 50 pilot symbols at rate $f_{pred} = 100$ kHz are utilized. While this pilot rate is too high when the objective is to overcome the feedback delay for the frequency-division duplex systems [8], it is sufficient in the half-duplex system that employs CTS CSI due to small delays $\tau_l$. We assume that the SNR of the pilot symbols is 30 dB. This effective pilot SNR can be achieved using noise reduction [18] when the actual SNR ≈ 20 dB, the value used in the network simulations in Section V. Assuming the carrier frequency 2.4 GHz and flat Rayleigh fading channel, Figure 5 shows the analytical mean-square-error (MSE) performance of the RTS CSI (outdated CSI without prediction) and CTS CSI method for a system with $L = 4$ relays. We observe that the proposed method significantly improves the CSI accuracy.

The proposed CTS-CSI method reduces overhead relative to [4] since it does not feed back CSI in the RTS frame. Realistic channel prediction involves adaptive linear filtering and insertion of low rate pilot symbols in the CTS frame and does not impose significant overhead or excessive computational load.

## V. NETWORK SIMULATIONS

Performance comparison is carried out using the OMNET++ network simulator [14] with the INETMANET package [19]. The transmit power and noise level are fixed at 1.0 mW and -102 dBm, respectively. At the physical layer, the channel consists of independent flat Rayleigh fading on each link with two-ray path loss as the large-scale model. Our implementation of Rayleigh fading utilizes the Clarke's model-based simulator [20]. For the path-loss model, the antenna heights are fixed at 1.5m, and the antenna gains are $G$ = 1 at both the transmitter and the receiver. Assuming the receiver sensitivity is -93 dBm, the approximate transmission range will be 300 m in the absence of fading at the data rate of 1 Mbps. We assume perfect estimation of the average channel power of the received HELLO messages for large-scale channel adaptivity.

The network topology consists of 50 mobile nodes in a square region of size 500m x 500m. The initial position of the nodes is random. Node mobility follows the Random Waypoint Mobility (RWP) model with node velocity uniformly distributed between $(0, v_{max})$, where $v_{max}$ is the maximum velocity, and a pause time of 2 seconds. Data traffic is generated by ten randomly selected sources, each producing 256 byte packets every 0.25 seconds to send to ten randomly selected destinations. The source-destination pairs are mutually exclusive, i.e. no node is the source or destination of more than one data connection. The start times for the data traffic flows are randomly distributed over the time interval of [10sec, 200 sec], and all sources stop transmission of data at 1000 seconds. Results are averaged over 50 independent simulation runs.

The average end-to-end throughput for the CTS CSI and RTS CSI methods is shown in Figure 6(a). We also plot the throughput for the ideal CSI method as the benchmark. The ideal CSI method does not require the RTS/CTS exchange as it assumes perfect CSI knowledge for all neighbors. We observe that the CST CSI outperforms the RTS CSI for all $L$. The gain of the CTS CSI scheme increases with $L$ and the mobile speed. Also, as L increases, the delay by which the channel measurement is outdated also increases. Hence, $L = 3$ outperforms $L = 4$ for both the RTS CSI and CTS CSI methods as the mobile speed increases. Note that $L = 2$ case is not shown in Figure 6(a), but it follows similar trends to other values of $L$ and has lower throughput than for $L = 3$ and 4.

In Figure 6(b), we plot the average end-to-end delay for the two methods. Note that the end-to-end delay grows with the number of relays. Also, the CTS CSI scheme has slightly larger delay than the RTS CSI scheme for higher velocities because it selects more robust routes, resulting in improved throughput. The end-to-end hop count follows a similar trend [**Error! Bookmark not defined.**].

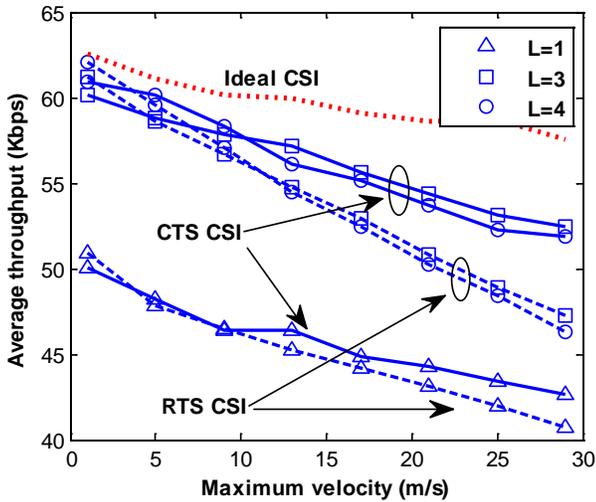

(a) Throughput

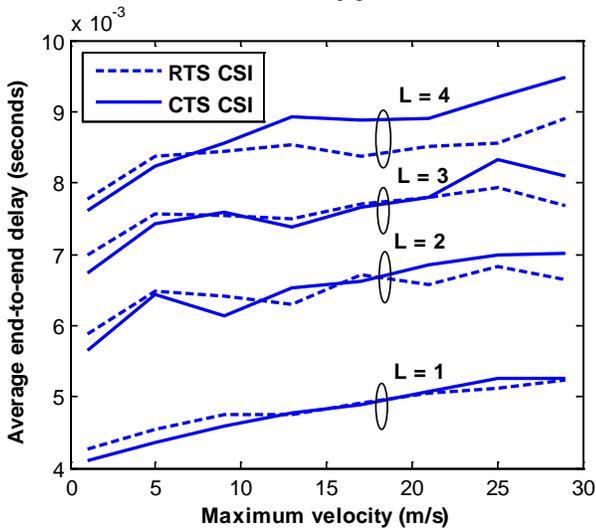

(b) End-to-end Delay

Figure 6: Comparison of RTS CSI (dashed lines) and CTS CSI (solid lines), Mobile topology, Rayleigh fading channel.

Finally, we observe that the rate-and-relay adaptive system achieves the best performance when the total number of relays is $L = 3$ since in this case the time delay associated with the CTS packet is lower than for $L = 4$, resulting in improved CSI accuracy. Thus, $L = 3$ achieves high throughput without compromising the end-to-end delay.

## VI. CONCLUSION

The effect of imperfect CSI on the rate-and-relay selection in ad hoc networks was analyzed and evaluated using network simulations. A novel method to reduce the CSI delay based on channel reciprocity and fading prediction was investigated for 802.11 networks. It was demonstrated that the proposed method improves significantly on the conventional approach that employs the outdated CSI. Moreover, the effect of the CSI accuracy on the choice of the total number of relays in high mobility scenarios was investigated. Future work will focus on the analysis for realistic fading models with shadowing and the development of adaptive transmission algorithms robust to imperfect CSI.